# Magnetic interactions and the puzzling absence of any Raman mode in EuTiO$_3$

P. Pappas[1], E. Liarokapis[1], M. Calamiotou[2], and A. Bussmann-Holder[3]

[1]Department of Physics, National Technical University of Athens, Athens 15780, Greece
[2]Section of Condensed Matter Physics, Physics Department, National and Kapodistrian University of Athens, GR-15784 Athens, Greece
[3]Max-Planck-Institute for Solid State Research, Heisenbergstr. 1, D-70569 Stuttgart, Germany

**Abstract**

Polycrystalline ceramic samples and a single crystal of EuTiO$_3$ have been investigated by Raman spectroscopy in the temperature range 80-300 K. Although synchrotron XRD data clearly indicated the cubic to tetragonal phase transition around 282 K, no mode from the symmetry allowed Raman active phonons was found in the tetragonal phase, contrary to the case of the homologous SrTiO$_3$. In order to study the evolution of this unique characteristic, ceramics of Eu$_x$Sr$_{1-x}$TiO$_3$ (x=0.03-1.0) characterized by synchrotron XRD for the structural phase transition have been also investigated by Raman spectroscopy, verifying the very strong influence on the Raman yield by Eu substitution. By applying an external magnetic field or alternatively hydrostatic pressure modes are activated in the Raman spectra. The temperature dependence of the main mode that is activated shows remarkable agreement with theoretical predictions. We attribute the puzzling absence of the Raman modes to a mechanism related to strong spin-lattice interaction that drives the cubic to tetragonal structural phase transition and makes the Raman tensor antisymmetric. On the contrary, the external perturbations induce a symmetric Raman tensor allowing even symmetry modes to be present in the spectra. Previous EPR, muon scattering and magnetic measurements indicated the presence of small magnetic interactions deep inside the paramagnetic phase. In order to probe those magnetic interactions in our EuTiO$_3$ polycrystalline sample and test our hypothesis, we have performed temperature dependant XAS/XMCD, which support the existence of magnetic nanodomains even close to room temperature.

## I. Introduction

Multiferroics are a growing field of scientific and technological interest, which has been in the focus of research for the last decade. Due to their feasibility to tune magnetic properties by an electric field and electric ones by a magnetic field, they are promising candidates in technological applications involving novel nano-electronics, spintronics, etc. Perovskite oxides (ABO$_3$) are well known for their various technological applications associated with their dielectric, optical, piezoelectric, and ferroelectric properties. An interesting subclass of those oxides comprises materials that exhibit magnetoelectric coupling between a soft optic mode and the magnetic moment of the ion occupying the A-site. A particular magnetoelectric perovskite titanate system, namely EuTiO$_3$ (ETO), has raised the interest of the scientific community since it is on the verge of multiferroicity. The presence of the magnetic 4f$^7$ Eu$^{+2}$ ions (S=7/2) induces a paramagnetic to G-Type antiferromagnetic phase transition at T$_N$=5.5K[1] and an abrupt 7% drop in the dielectric constant at T$_N$ that indicates strong magnetoelectric coupling. The temperature dependence of the dielectric constant has been related to a long-wave-length soft transverse optic mode suggesting polar properties of ETO.[2] In analogy to SrTiO$_3$ (STO) the transition to the ferroelectric phase is suppressed,[3] however with an extrapolated transition temperature of -150K[3], whereas in STO this lies at 35.5K.[4] Although temperature does not act as the driving force for a ferroelectric transition of ETO, strain[5] and electric field[6] have been shown to induce possible polar properties in thin films of ETO. For STO it has been demonstrated that an external field can induce a long range polar phase.[7,8] Furthermore, for STO a ferroelectric transition takes place upon isotopic O$^{16}$/O$^{18}$ substitution or by replacing Sr with small amounts of Ca.[9,10] For ETO short range magnetic fluctuations at high temperatures, well above T$_N$,[11,12] have been detected and the expected coupling with local dipole moments associated with the Ti$^{+4}$ ion revives the possibility for ETO to be a truly multiferroic material.

At ambient conditions ETO crystallizes in the same cubic structure ($Pm\bar{3}m$) as STO with almost the same lattice constants (a=3.9050Å [1] for STO and a=3.9082Å[13],a=3.9058Å[14] for ETO).[15,16] In the cubic structure three IR-active modes with F$_{1u}$ symmetry have been observed[2] none being Raman active. From the three transverse optic IR active modes the first one (TO1) is the soft mode and its eigenvector is associated with a predominantly Slater type mode. The other two IR active modes are, a predominately Last mode (TO2), which involves the motion of the TiO$_2$ cage against Eu atoms and the Axe mode (TO4),[17] which displays the bending of the oxygen octahedra. Theoretical and several experimental results have shown that an antiferrodistortive structural phase transition to a tetragonal phase takes place in ETO at T$_s$≈282 K,[13,14,16] analogous to the one of STO at $T_s$=110 K.[15] In the tetragonal phase (space group *I4/mcm*) factor group analysis yields seven Raman active modes. For STO extensive Raman measurements including the application of an electric field [7,15] verified the structural change from cubic to tetragonal or orthorhombic crystal symmetry.[7] In this work we investigate the complete absence of any Raman active mode in the tetragonal low temperature phase of ETO contrary to our and other diffraction measurements that definitely show the phase transition to tetragonal symmetry.[13,14] Furthermore, we show for the first time that modes can be activated in Raman scattering by applying an external dc magnetic field or by using hydrostatic pressure. A plausible explanation of the observed effects is also provided based on a phenomenological model.

## II. Experimental Setup

For the XRD and Raman measurements $Eu_xSr_{1-x}TiO_3$ (x=0.03-1.0) polycrystalline samples have been used. The polycrystalline samples were synthesized with solid state reaction methods at the Max-Planck-Insitut für Festkörperforschung in Stuttgart.[16] Also a small single crystal of pure ETO synthesized at Cambridge University by the floating zone method[18] and provided by Christos Panagopoulos was studied. More details about the synthesis of the samples have been presented elsewhere.[1,16,18]

Raman measurements, in backscattering geometry, have been carried out with a T64000 Jobin-Yvon triple spectrometer equipped with a CCD detector using an Ar+ laser with the 514.5 nm line and a solid state laser with 532 nm line at low intensity (0.1-0.5 mW) to avoid heating of the sample. Low temperature measurements were performed with a LINKHAM and an Oxford Instruments low temperature cryostats. Measurements under hydrostatic pressure up to 6 GPa have been carried out using a Merrill-Bassett type diamond anvil cell (DAC) and a 4:1 methanol ethanol mixture as pressure transmitting medium, while a GaP single crystal was used as a pressure calibrator. For the low temperature high pressure measurements the DAC was placed inside the Oxford cryostat operating in the range 80-300 K. Raman scattering measurements with in situ application of a magnetic field were performed by placing appropriate Nd magnets under the sample inside the cryostat. The magnetic field on the surface of the sample was up to 0.4 T.

Additional UV Raman measurements with a single stage Princeton Instruments spectrometer have been performed at IUVS beamline at Elettra Sincrotrone Trieste using a 266 nm solid state laser. A series of Raman spectra were acquired at temperatures ranging from 80K to 300 K with a LINKHAM cryostat that agreed with those in the visible spectrum.

XAS/XMCD measurements were performed on the polycrystalline ETO sample in Total Electron Yield (TEY) detection mode at the synchrotron Solaris in Krakow in order to probe any local magnetic field. We have chosen the Eu $M_{4,5}$ (1155/1128 eV) absorption edge for data collection, which corresponds to the atomic transition $3d^{10}4f^7 \rightarrow 3d^9 4f^8$. By probing the $M_{4,5}$ edge, information about the $4f^7$ ground state involved in the magnetic interactions was obtained. The analysis of the XAS/XMCD spectra was performed with the PyMca software.[19] Due to the strong spin-orbit coupling in both the ground and the excited state, the number of possible atomic transitions is large especially for this half filled f shell. Also, multiplet interactions play a significant role in the rare earth $M_{4,5}$ edges. On the other hand, the localized nature of the 4f electrons makes them unaffected by the local environment and the XAS/XMCD spectra can be interpreted by the atomic multiplet theory.[20] For the application of the magnetic field the in-house liquid nitrogen cooled electromagnet was used, which could deliver up to 0.2 T in the same direction as the incident light. The sample was placed on a cold finger inside the UHV chamber and the probed temperature range was 30-300 K. Taking advantage of XMCDs high orbital selectivity the presence and temperature dependence of small local magnetization was detected from the integrated intensity of the XMCD signal, which is proportional to the magnetic moment of the absorbing atom. The characterization of our $EuTiO_3$ polycrystalline sample was carried out at the beamline XRD1 of the Elettra Sincrotrone at Trieste, using the 2M PILATUS detector. Powder diffraction data have been collected with $\lambda=0.59043$ Å wavelength in the Debye Scherrer geometry. The characterization of the $Eu_xSr_{1-x}TiO_3$ x={1.0,0.75,0.5,0.25,0.03} polycrystalline mixed crystals was carried out at the MCX High resolution beamline of the Elettra Sincrotrone at Trieste. The wavelength used was $\lambda=0.61946$ Å in the Debye Scherrer geometry. For the low temperature measurements at both beamlines an Oxford Instruments cryostream cooler was used. Temperature calibration was performed with $LaB_6$ at XRD1 and with NaCl at MCX beamline. Diffractograms were indexed by using the DICVOL91 software[21] and Rietveld refinements were performed by using the Fullprof suite.[22] In particular, for the Rietveld refinements the background was fitted by using a linear interpolation between a set of given background points. In the last refinement cycle scale factor, zero correction, profile share parameters, positional coordinates (if allowed by site symmetry) and cell parameters were allowed to vary until convergence was reached.

### III. Experimental results
a) XRD characterization

Data acquisition for $EuTiO_3$ revealed that there was no extra peak associated with parasitic phases for all polycrystalline samples. By indexing the low temperature XRD patterns the best fit was, as expected, a tetragonal cell with lattice parameters a=b=5.5085 Å and c=7.7660 Å with the systematic extinctions pointing towards a body centered cell. As previous work has pointed out,[23] the most probable space group for the low temperature phase is the *I4/mcm* tetragonal space group with $\sqrt{2}a \times \sqrt{2}a \times 2a$ unit cell metric, where *a* is the primitive cubic cell lattice parameter. The tetragonal structure in the *I4/mcm* space group involves an out-of-phase tilting of the oxygen octahedra around the tetragonal axis associated with the $R_4^+$ irreducible representation.

Structural information for the polycrystalline samples was acquired by Rietveld refinement. Profile matching for pure EuTiO$_3$ and the mixed Eu$_x$Sr$_{1-x}$TiO$_3$ samples at ambient conditions yielded good agreement with the $Pm\overline{3}m$ cubic space group and at low temperatures with the *I4/mcm* one (Fig.1). Our refined lattice parameters for ETO (*a*=3.90498 Å for XRD1 data and 3.90337 Å from MCX data) are in agreement with previously published results.[13,23] The evolution of the (400)$_{cubic}$ peak splitting with temperature (inset in Figure 1) points to a T$_S$ close to 300 K for EuTiO$_3$ as expected.[13,14] By following the splitting or the broadening of the (400)$_{cubic}$ reflection, which is expected to split in the tetragonal phase into (008)$_{Tetragonal}$ and (440)$_{Tetragonal}$ at low temperatures, the existence of the expected antiferrodistortive structural phase transition below T$_S$ for the whole series of Eu$_x$Sr$_{1-x}$TiO$_3$ samples was verified. The transition temperatures estimated for all Eu concentrations by plotting the FWHM of the (400)$_{cubic}$ reflection as a function of temperature agree with those from EPR measurements[24] (Figure 2).

b) Low temperature Raman measurements

At ambient conditions STO and ETO have the same cubic symmetry (space group $Pm\overline{3}m$) with three T$_{1u}$ IR active and one T$_{2u}$ silent mode. In the low temperature (LT) phase (below 284 K, space group *I4/mcm*) factor group analysis[25] shows that ETO (and STO) has 5E$_u$+3A$_{2u}$ IR active modes and A$_{1g}$+3E$_g$+B$_{1g}$+2B$_{2g}$ Raman active ones. The relation between the irreducible representations of the modes after the Brillouin zone folding due to the cubic to tetragonal (C-T) symmetry change is presented in Table I. Despite the fact that our synchrotron XRD data confirm such a C-T phase transition for the EuTiO$_3$ sample close to room temperature (RT), no sign of any of the expected modes appears in the Raman spectra down to liquid nitrogen temperature, with both visible (bottom spectrum in Figure 3) and UV excitation sources. As Eu$^{+2}$ substitutes for Sr$^{+2}$ at small amounts (x=0.03) the Raman spectra are reminiscent of STO[7] except for the 172 cm$^{-1}$ band, possibly an impurity. At x=0.25 and higher Eu concentrations the Raman yield is substantially reduced indicating the very strong influence that Eu has on Raman scattering. It should be noted a weak broad band appears around 500 cm$^{-1}$, which is absent in pure EuTiO$_3$ at room or low temperatures (Figure 3). The apparent differences between STO and ETO, i.e., broad bands above and narrow ones below T$_S$ for STO and complete absence of both at all temperatures in ETO must originate from differences between $Sr^{+2}$ and $Eu^{+2}$ ions, since the structure and the lattice constants are almost identical. From the measurements described in the following we conclude that the origin of the discrepancy lies in the magnetic properties of the $Eu^{+2}$ ions.

Recently, an experimental study of ETO polycrystalline samples[26] revealed broad Raman bands even at room temperature (as in $SrTiO_3$, Refs.[7,8]), which were almost temperature independent. Based on the information provided in Ref.[26] we suspect that the broad bands are the result of the high laser power used in these studies. Figure 4 clearly evidences the effect of overheating on our samples: starting from a featureless Raman spectrum, consistent with the cubic space group, broad bands appear with increased laser power, which are similar to those observed in Ref.[26]. From this observation we conclude that the observed modes detected[26] are caused by the dissociation of the ETO, possibly to titanium oxide.

c) High pressure Raman data

Figure 5 presents characteristic hydrostatic pressure spectra at RT for a single crystal of ETO. It is found that for pressures greater than ≈4.1 GPa a mode ~60 cm$^{-1}$ is activated in the Raman spectra, which disappears when pressure is decreased below 4.1 GPa. With further increase of the hydrostatic pressure the mode clearly hardens and at higher pressures another mode at ≈600 cm$^{-1}$ starts to develop (Figure 5). The observed modes cannot be due to a reduction in symmetry from anisotropic strains since for such small pressures the transmitting medium (methanol-ethanol) is expected to be hydrostatic[27]. The possibility of sample inhomogeneity was investigated by acquiring Raman spectra from various spots at 3.2 GPa and 4.2 GPa. In all cases for 3.2 GPa the Raman spectra were featureless, while at 4.2 GPa the low energy mode was present but with energy varying in the range ≈70-113 cm$^{-1}$ depending on the spot.

As the temperature is lowered the pressure-induced mode hardens with decreasing temperature (top spectra in Figure 5), while the threshold pressure for activation is lowered from roughly 4.1 GPa at 300 K to ~3.2 GPa at 250 K. The difference cannot be attributed to the slight contraction of the anvil cell, which is negligible for such small changes in temperature. Complementary hydrostatic pressure measurements under the same conditions of the polycrystalline ETO have shown the same behavior although the threshold was lower for the polycrystalline sample (≈3.7 GPa at RT) and the low energy mode was observed at ≈61 cm$^{-1}$. The existence of a threshold pressure has already been established by synchrotron XRD measurements.[28,29] Furthermore, it was found that the critical pressure increases with temperature[28] in agreement with the Raman results.

d) Magnetic Field induced Raman scattering

Figure 6 presents the phonon activation effect in the Raman spectra from a small dc magnetic field (~0.2 T) on polycrystalline ETO. For this magnetic field a mode appears in the low energy region (~40 cm$^{-1}$) for temperatures below ~250 K, i.e., at temperatures well above the Néel temperature. As the temperature is reduced, the new mode hardens substantially (Fig. 6). The exact temperature for the disappearance of the mode is uncertain since with increasing temperature the mode becomes wider and dives into the strong elastic scattering continuum of the Raman spectra. From theoretical calculations[30] a comparable temperature dependence of the transverse zone boundary mode has been deduced which is shown in Fig. 7. The zone boundary mode has been also measured by inelastic x-ray scattering.[31] The agreement between the measured mode frequency under two small magnetic fields and the theoretical predictions is remarkable showing that the activated mode is the transverse zone boundary mode. This effect is intimately related with the findings from birefringence, EPR and muon spin rotation (μSR) experiment where a small magnetic field activated birefringence in a EuTiO$_3$ thin film[32] and magnetic activity was detected.[12] Also intrinsic magnetism even in the paramagnetic phase due to strong spin lattice coupling is predicted [12][30] since the second nearest neighbor exchange constant directly depends on the spin-lattice coupling constant. The mode that is induced by the magnetic field resembles the one presented above 4.1 GPa hydrostatic pressure and also the temperature dependences of the two modes are similar. As indicated by the data points in Fig.7. at low temperatures for 4.2 GPa pressure and at RT with increasing pressure the frequency of the mode shifts to higher values. Removal and re-application of the magnetic field revealed that strong hysteresis phenomena are present. When the sample is field cooled the removal of the magnetic field leaves a permanent magnetic effect for several hours. Also, the emerged phonon mode remains as long as the sample is magnetized showing the close relation between the activated phonon mode and the local magnetic field.

e) XAS/XMCD results

Many studies on rare earth compounds have pointed out the tendency of the rare earth ion to fluctuate between +2 and +3 valence state depending on their position in the lanthanide group, their chemical environment, and synthesis conditions.[33,34] By comparing the simulated XAS spectrum for both Eu$^{+2}$ and Eu$^{+3}$ oxidation states based on the CTM4XAS software[35] with the experimental ones (Figure.8.) it turns out that there is a mixed valence state at the very surface (few nanometers) of our sample with almost equal contribution of both Eu valences. Recently, a detailed study about the synthesis of ETO samples with Eu$^{+2}$/Eu$^{+3}$ mixed valence state has revealed valuable information about the structural properties of those compounds.[36] By comparing the very small parasitic peaks in Figure.8. (inset) of our synchrotron XRD results with those in Ref.[36] (Fig.2b) we conclude that the contribution of the $Eu^{+2}$ valence dominates the bulk of our polycrystalline sample. Furthermore, the Raman spectra presented in Ref.[36] fully agree with our observations of the complete absence of any peak in pure ETO, while in the mixed valence samples distinct peaks appear,[36] which are absent in our Raman spectra confirming that in the bulk of our polycrystalline ETO sample any presence of $Eu^{+3}$ is very limited and we have a clear domination of $Eu^{+2}$. Furthermore in Ref.[36] it is pointed out that extreme caution should be exercised with respect to the laser power used in Raman measurements, because heating the sample in open air can act as an oxidizing atmosphere thus affecting the stability of Eu valence. This confirms our findings presented in Figure 4 regarding the effect of increased laser power to the Raman spectrum.

The XMCD spectra for both $M_{5,4}$ edges at different temperatures is shown in Figure 9. Without an external magnetic field, the XMCD signal increases substantially below 200 K while by applying 200 mT field it passes from a maximum around 70 K. It is interesting that a none zero XMCD signal exists even at 290 K without an external magnetic field, but this is possibly related with a small remaining magnetization from previous Raman measurements under an external magnetic field due to hysteresis effects. By applying specific sum-rules the separation of the total magnetization into an orbital and a spin part can be obtained.[37,38] The temperature dependence of the spin part of the magnetization (namely $m_S$) and the effect that an external magnetic field has on it were obtained from the integration of the XMCD/XAS spectra (M5, M4 edges) using the PyMca software.[19] The relative amount of each contribution was based on the similar analysis followed to other rare earths compounds.[39,40] As seen in the inset of Figure 9, without an external magnetic field the $m_S$ is found to increase mainly below ~200 K, to reach 0.6 µB at 30 K. Qualitatively the temperature dependence of $m_s$ resembles the evolution of macroscopic magnetization (Figure 2 in Ref.[12]) although here an external magnetic field is not present and the maximum $m_s$ corresponds to 8.5% of the maximum value for a fully spin polarized case (7 $\mu_B$). Possibly the increase at ~200 K (with or without an external magnetic field) and at ~70 K (with the field) are related with those observed in Ref.[12] of muon scattering measurements and in the birefringence of ETO epitaxial film.[32]

## IV. Discussion

Despite its cubic symmetry that forbids Raman active modes, several broad bands appear at ambient conditions in $SrTiO_3$, attributed to the large anisotropy of the oxygen polarizability.[41] EXAFS measurements on two oxygen isotopes of STO revealed two distinct groups of oxygen atoms separated by 0.1 Å, that survive to temperature much higher to Ts and the effect is more pronounced in the $^{18}O$ samples.[42] The second coordination of Sr-Ti indicates also two interatomic distances, which resembles the situation in $BaTiO_3$.[43] Similar distortions have been observed in STO,[44] which can explain the broad bands observed even at room temperatures in the Raman spectra.[41] At low temperatures, in the tetragonal phase, narrow peaks appear as expected in agreement with the differences in the local structure.[43,44] On the contrary, for ETO no modes or broad bands appear in the Raman spectra at room or low temperatures in all our samples. In terms of local structure, ETO appears to be different than STO; Eu-Ti distance does not show the pronounced double peak of STO at low or high temperatures,[43] apparently indicating less oxygen disorder in ETO. This would justify the absence of broad bands at RT, but cannot explain the absence at low temperatures in the tetragonal phase. From the few data that exist for the local structure of ETO, it appears that oxygen atoms occupy two different Eu-O positions, which are not modified appreciably in the whole range 10-280 K.[43] In STO the relative Sr-O distances are strongly distorted close to RT with at least three atomic distances to gradually settle down to one position below $T_s$.[43] This icon agrees with the strong oxygen disorder associated with the broad Raman bands in STO (or in low Eu concentration of $Eu_{1-x}Sr_xTiO_3$, Fig.3) and the narrow peaks in the tetragonal phase.[8] Furthermore, it comforts with the findings that the double well potential is shallow and wide for STO and deep and narrow in ETO[16]. The puzzle is the complete absence of any mode in the Raman spectra at low temperatures. In the following, we examine some reasonable assumptions that could explain the surprising complete absence of such modes in the tetragonal phase, where they are allowed by symmetry, as well as the origin of the modes appearing under the action of an external magnetic field.

One possibility could be the structural phase transition to a lower symmetry group that does not allow Raman active modes. By indexing the synchrotron XRD data of the samples used in this work, but also almost all relevant studies such as Ref.[23] clearly reveal that a tetragonal $\sqrt{2}a \times \sqrt{2}a \times 2a$ unit cell can better reproduce the diffraction pattern. All the possible other subgroups of $Pm\bar{3}m$ that involve tilting of the oxygen octahedra[45] seem to have Raman active modes as well. To the best of our knowledge only one experimental study suggests the absence of an antiferrodistortive structural phase transition close to RT proposing a scheme with Eu ions delocalization and short range coexistence of different crystallographic phases.[46] But, even in that scheme Raman active modes cannot be excluded by symmetry.

Another option is that Raman active modes indeed exist, but they are strongly overdamped due to disorder, mainly at the oxygen ion sites, since Raman modes eigenvectors involve mostly the oxygen ions. On the average, such disorder should result in a small antiferrodistortive oxygen octahedral rotation, in agreement with the observed doubling of the unit cell at low temperatures. Already published experimental studies have revealed that intrinsic disorder in ETO is present which also affects the structural phase transition.[23] But, as is the case of SrTiO$_3$, disorder at the oxygen ion sites would induce broad bands even above $T_s$, since the $q \approx 0$ selection rule is violated. In addition, the possible strong disorder should show up in the width of the powder diffraction lines in contrast to the observations. On the other hand, EXAFS data show that oxygen disorder is less in ETO than in STO,[43,44] explaining the absence of broad bands in the cubic phase of ETO.

The fact that a moderate magnetic field induces Raman modes at low temperature and the correlation between the remaining magnetism and the activated mode demonstrates an effect related to the magnetic properties of the compound. The reminiscent magnetization after the removal of the external magnetic field is a clear indication of the strain induced by the strong spin lattice interaction inside the nanodomains. Although ETO is in the paramagnetic phase the presence of those magnetic but randomly oriented in space nanodomains resembles the behavior of relaxor ferroelectrics. Evidence of dynamic magnetism at elevated temperatures and enhanced spin-lattice coupling is available from magnetic, muon, and EPR measurements[11,12] indicating that around ~200 K there is some kind of magnetic crossover temperature. Our XMCD measurements revealed the existence of local magnetic moments that survive close to room temperature, in agreement with EPR and muon scattering measurements,[11,12] and indicate two characteristic temperatures around 70 and 200 K. Also a clear effect of a magnetic field on the structural phase transition temperature[12] and the corresponding acoustic soft mode[11] proves that the spin fluctuations are strongly correlated with the lattice dynamics. Birefringence data clearly support that a magnetic field of the same order of magnitude as the magnetic fields used here affects the structural phase transition temperature.[32]

As discussed before, pressure measurements have shown that beyond a critical hydrostatic pressure (~4.1 GPa) a mode is activated in Raman scattering similar to the one induced by the external magnetic field. The threshold pressure is in agreement with previous XRD measurements.[28,29] This observation can be understood in terms of decreasing the Eu-Eu and Eu-O-Eu distances with pressure, which modify the exchange coupling constants in agreement with magnetostriction data. Also, the substitution of $Eu^{+2}$ for $Sr^{+2}$ in the $Eu_xSr_{1-x}TiO_3$ which modifies the average Eu-Eu distances, supports this interpretation. It is therefore reasonable to assume that the absence of Raman active modes is caused by the "hidden" magnetic properties of the compound.

In the presence of an external magnetic field the Raman tensor will acquire an antisymmetric component that can only couple to certain symmetry modes.[47] This is due to the breaking of time reversal symmetry of the magnetic field, which makes the eigenfunctions of the system not real in general. The perturbation causes a reduction of the crystal symmetry activating or splitting some degenerate phonon modes. The same effect will happen if the magnetic field is internal. Accordingly, the Hamiltonian can split into a real part that corresponds to the unperturbed cubic system and an imaginary one, linear (in the first order) to the (internal) magnetic field:[47]

$$H(B) = H^{(0)} + iH^{(1)}(M) \qquad (1)$$

where both $H^{(0)}$ and $H^{(1)}(M)$ are real and $M$ defines the local magnetization from the internal field. The electronic polarizability also will decompose into a real symmetric part $\alpha^{(0)}_{\mu\lambda\sigma}$ of the original unperturbed cubic symmetry crystal and a term induced by the magnetic field $\alpha^{(1)}_{\mu\lambda\sigma}(M)$, which is neither symmetric nor real,[47] and, correspondingly, the Raman tensor can be expressed as:[47]

$$\begin{aligned}\alpha_{\mu\lambda\sigma}(M) &= \alpha^{(0)}_{\mu\lambda\sigma} + \alpha^{(1)}_{\mu\lambda\sigma}(M) \\ &= \alpha^{(0)}_{\mu\lambda\sigma} + ib_{\mu\lambda\sigma\nu}M_\nu\end{aligned} \qquad (2)$$

Assuming that the $\alpha^{(0)}_{\mu\lambda\sigma}$ term of EuTiO$_3$ corresponds to the cubic phase, where no Raman active modes are allowed, the tetragonal phase will be induced by the internal magnetic interactions, which activate the $\alpha^{(1)}_{\mu\lambda\sigma}(M)$ term. The presence of dynamic magnetism in nanoregions[12] could be the origin of such an internal magnetic field. Accordingly, the Raman tensor will consist of the symmetric term stemming from the cubic phase and an asymmetric one that is associated with the internal field and the tetragonal phase. For the D$_{4h}$ point group the Kronecker product of the irreducible representation of electric moment is:

$$\Gamma(M) \times \Gamma(M) = 2A_{1g} + B_{1g} + B_{2g} + 2E_g + A_{2g} \qquad (3)$$

The pure antisymmetric part of the Kronecker product is $[\Gamma(M) \times \Gamma(M)]_{asym} = A_{2g}$. In the case of D$_{4h}$ point group and an asymmetric Raman tensor, only the A$_{2g}$ modes are Raman active. For EuTiO$_3$ the expected Raman active modes (Table I) do not include phonons with A$_{2g}$ symmetry. Therefore, no mode is allowed by symmetry, provided the Raman tensor is antisymmetric. XRD and IR measurements are independent of this effect and therefore will detect the structural phase transition. Other titanates with rare earth ions occupying the A-site of the perovskites lattice such as GdTiO$_3$, SmTiO$_3$ and YTiO$_3$ could in princible present similar effects. Keeping in mind that at ambient conditions ETO is the only rare earth titanate that crystallizes in the cubic Pm-m space group with no Raman active modes makes it a perfect system to illustrate the absence of any Raman active mode at low temperatures. Other rare earth titanates being already in a low symmetry phase with allowed Raman active modes, will not present such an anomaly due to local magnetic interactions. In those systems the magnetic field would only induce some shift or split of the Raman peaks, but it could not make them disapear. On the other hand, as pointed out in Ref.[47] for T>T$_S$ if there are no Raman modes, a magnetic field cannot induce them, that is why for ETO at room temperature the local magnetic interaction have no effect on the Raman spectrum. When an external magnetic field is applied, it couples to the local magnetic moments and gives rise to a term that respects time-reversal symmetry creating a symmetric Raman tensor component that can activate the observed mode (Figures.6.).

The increased transition temperature of ETO compared with STO could be explained by the polarizability model that incorporates the spin-spin and spin-phonon interaction.[16] The additional assumption made here is the local breaking of time reversal symmetry that activates certain mode in the tetragonal phase inaccessible by Raman spectroscopy. The exact form of the term that breaks time-reversal symmetry and at the same time respects the space inversion (otherwise, the IR modes will be "seen" in the Raman spectra), is uncertain. It could be an appropriate combination of terms of magnetization M, gradient $\nabla$, electric or magnetic toroidal moment, and polarization.[48] It is probably a magnetoelectric term that breaks time reversal symmetry and induces the doubling of the unit cell,[49] substantially increasing the transition temperature. In the absence of conclusive measurements on the lattice respond to a magnetic field, it is difficult to finalize the form of the interaction Hamiltonian. But, the clear evidence of a strong coupling between spin and lattice even above the structural transition temperature,[11,12] supports the theoretical modeling that this coupling enhances the second-nearest neighbor ferromagnetic interaction at a local scale.[30]

Concerning the effect of hydrostatic pressure, it is expected to modify the coupling of the spins and the spins with the lattice. If we ignore the effect of an external magnetic field, the 2$^{nd}$ order structural phase transition can be expressed by the Hamiltonian $H_{pt} = a(T,p)\xi^2 + b\xi^4$ where $\xi$ defines the order parameter (e.g., the rotation angle of the octahedra) and $p$ is the hydrostatic pressure. The coefficient $a$ will be given by $a(T,p) = a_o(T-T_o) - c_1 f(M) - \left[ c_2(p-p_o) + c_3(p-p_o)^2 \right]$ and $a_o$, $b$, $c_1$, $c_2$, $c_3 > 0$. $T_o$ defines the transition temperature in the absence of the internal field and can be much lower than RT, as it is the case for the non-magnetic STO and the $Eu_xSr_{1-x}TiO_3$ (Figure 2). We make the assumption that the second term is linear in the internal magnetic field (expressed via the magnetization M) or more generally it includes only terms that lead to the breaking of time-reversal symmetry. We assume a 2$^{nd}$ order dependence on the hydrostatic pressure based on the XRD measurements,[29] but higher order terms can be included without affecting our conclusions since they will induce a symmetric Raman tensor. For the value of $p_o$ we can assume a value on the order of 2-3 GPa.[28,29] It is possible that terms combining internal magnetization and hydrostatic pressure exist, but they will not alter the explanation concerning the Raman results. Both the effects of the internal local magnetization and the external hydrostatic pressure will tend to increase the transition temperature as compared to $T_o$. Since the term that induces the transition can be assumed linear in the (internal) magnetic field, the induced Raman tensor will be antisymmetric and the analysis presented above justifies the complete absence of modes in the Raman spectra. By applying a hydrostatic pressure, the transition temperature is increased and reaches RT at a critical hydrostatic pressure corresponding to ~4.1 GPa. At the same time, it will induce a time reversal symmetric term in the Raman tensor (similar with the one created by an external magnetic field), that causes modes to appear in the Raman spectra. The application of an external field will couple with the internal one giving a time-reversal term that activates the Raman modes in the tetragonal phase. The structural transition temperature is not modified appreciably by the applied external magnetic field (Fig.7.), but the fields employed in our measurements are not strong enough for final conclusions. However, such a simple model fully explains the Raman measurements and the strong hysteresis effect, assuming the existence of internal magnetic field regions, which are anticipated from the results obtained from μSR and EPR.[11,12] It must be pointed out that the magnetic nano-regions should be large enough so that the order parameter of the $2^{nd}$ order phase transition can be defined while the average magnetic effects should be minimal.

## Conclusions

Systematic Raman measurements have been carried out on a series of samples of pure $EuTiO_3$ and $Eu_xSr_{1-x}TiO_3$ mixed crystals at low temperatures and under the influence of an external magnetic field or hydrostatic pressure. The cubic to tetragonal structural phase transition temperature has been investigated by synchrotron XRD measurements and agrees with previously published results. The Raman studies evidence that in the tetragonal phase of $EuTiO_3$ all symmetry allowed modes are absent. The application of an external magnetic field or hydrostatic pressure activates some modes with remarkable agreement with theoretical predictions for the mode indicative of the structural phase transition. We attribute the observed phenomena as originating from the magnetic Eu ions, which create domains with a local magnetic field and enhance the structural phase transition close to RT, much larger than in the homologous $SrTiO_3$. By assuming an antisymmetric Raman tensor induced by the local field, we can explain the absence of Raman active modes, as well as the activation of modes by the external perturbations. By probing the local magnetic domains using the XMCD signal, we could verify the existence of such local magnetic fields close to room temperature.

## Acknowledgements

We thank J. Köhler of Max Planck Institute for providing the polycrystalline samples and Christos Panagopoulos from Cambridge University for the single crystal used in these studies, Elettra Sincrotrone in Trieste for beam time allocation and the staff of beamlines XRD1, MCX, and IUVS for their valuable assistance and discussions during the experiments. Also we would like to thank

SOLARIS synchrotron at Krakow for beam time allocation and the beamline staff for valuable support during our XAS/XMCD experiments and useful comments for data analysis.

This research is co-financed by Greece and the European Union (European Social Fund- ESF) through the Operational Programme «Human Resources Development, Education and Lifelong Learning» in the context of the project "Strengthening Human Resources Research Potential via Doctorate Research" (MIS-5000432), implemented by the State Scholarships Foundation (IKY). The experiment at the IUVS beamlineand Solaris was funded by CERIC-ERIC (#20192051).

We acknowledge fruitful discussions with H. Keller and K. Roleder.

Table I. Relations between the high and low symmetry irreducible representations.

| k-vector | Modes Irreducible Representation | |
|---|---|---|
| | $Pm\bar{3}m$ ($O_h$ point group) | $I4/mcm$ ($D_{4h}$ point group) |
| $\Gamma(0,0,0)$ | $T_{1u}$ | $A_{2u}$ |
| | | $E_u$ |
| | $T_{2u}$ | $B_{2u}$ |
| | | $E_u$ |
| $R(-1/2,1/2,1/2)$ | $R^+_4$ | $B_{1g}$ |
| | | $E_g$ |
| | $R^+_3$ | $A_{1g}$ |
| | | $B_{2g}$ |
| | $R^+_1$ | $B_{2g}$ |
| | $R^-_5$ | $A_{2u}$ |
| | | $E_u$ |

**Figure captions**

FIG. 1. Experimental (circles, red in color) and calculated (black solid line) diffraction intensities of EuTiO$_3$ at 80 K after Rietveld analysis. Vertical bars indicate the Bragg positions of tetragonal symmetry. In the inset the splitting of the (400)$_{cubic}$ reflection is depicted.

FIG. 2. Evolution of T$_s$ of Eu$_x$Sr$_{1-x}$TiO$_3$ with x compared with published EPR results[24].

FIG. 3. Typical Raman spectra of Eu$_x$Sr$_{1-x}$TiO$_3$ at ambient conditions and of EuTiO$_3$ at 80 K.

FIG. 4. Thermal effect of laser intensity on the EuTiO$_3$ ceramics.

FIG. 5. Single crystal Raman spectra under hydrostatic pressure at room temperature and their evolution with temperature.

FIG. 6. Raman spectra of EuTiO$_3$ under 200 mT DC magnetic field and after field cooling and removal of the small magnetic field (hysteresis effect).

FIG. 7. Zone Boundary acoustic mode temperature dependence for two magnetic fields (200 and 400 mT) compared with theoretical predictions.[30] The shift of the mode with hydrostatic pressure at room and low temperatures is also presented. Dashed line is just guide to the eye.

FIG.8. Comparison of experimental XAS spectrum at 290 K with simulated spectra of pure Eu$^{+2}$/Eu$^{+3}$ valence state and mixed valence state. Inset shows parasitic peaks of ETO XRD pattern at 300 K. The diffractogramm is normalized with respect to the intensity of the central peak.

FIG.9. XMCD spectra of Eu M$_{5,4}$ absorption edges at various temperatures with 0 mT and 200 mT external magnetic field. Inset shows the temperature dependence of spin contribution, m$_S$, to the total magnetization. Lines are just guide to the eye.

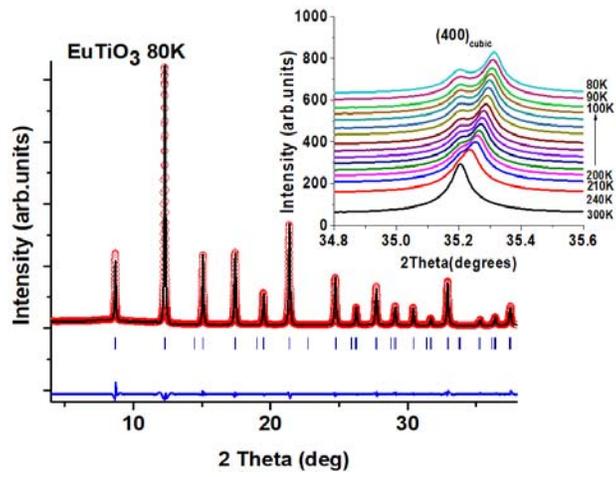

FIG. 1.

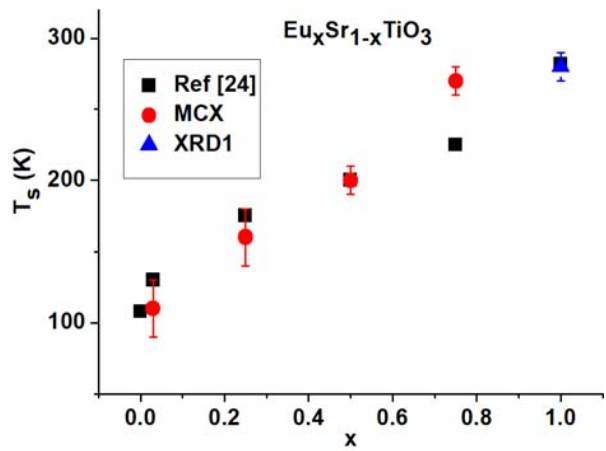

FIG. 2.

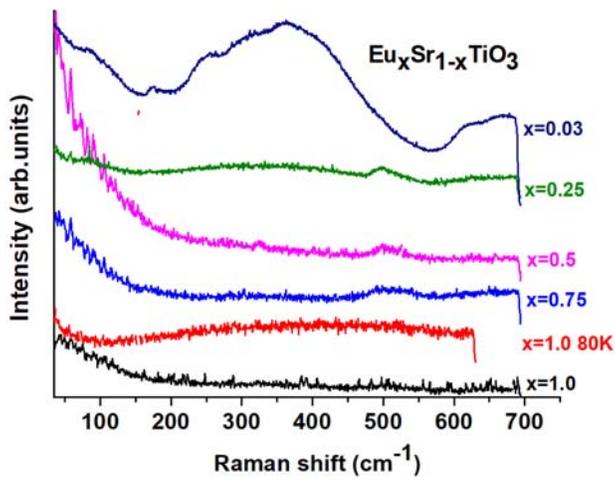

FIG. 3.

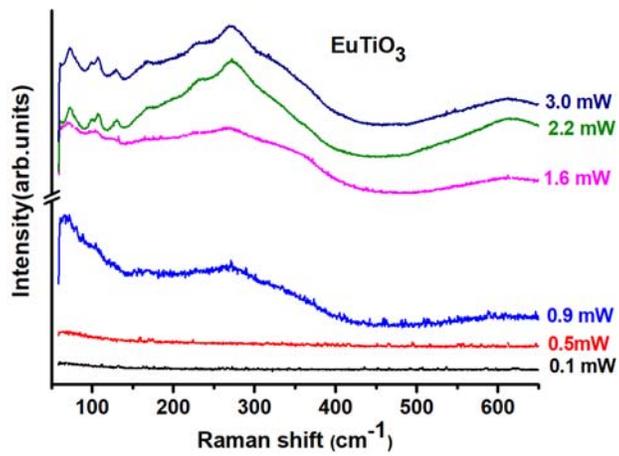

FIG. 4.

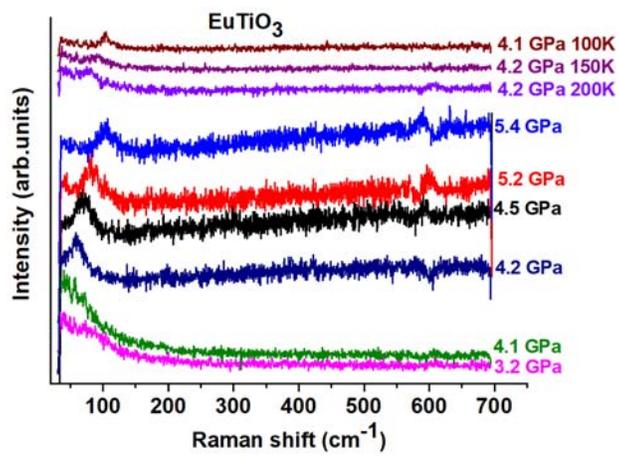

FIG. 5.

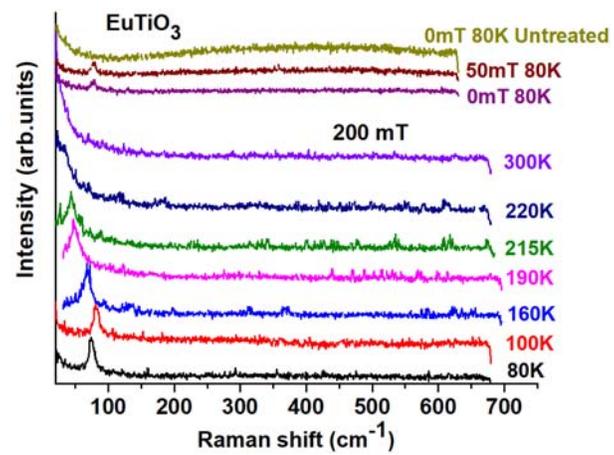

FIG. 6.

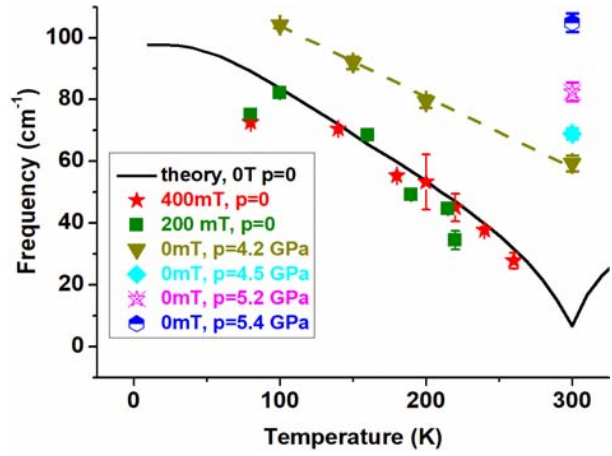

FIG. 7.

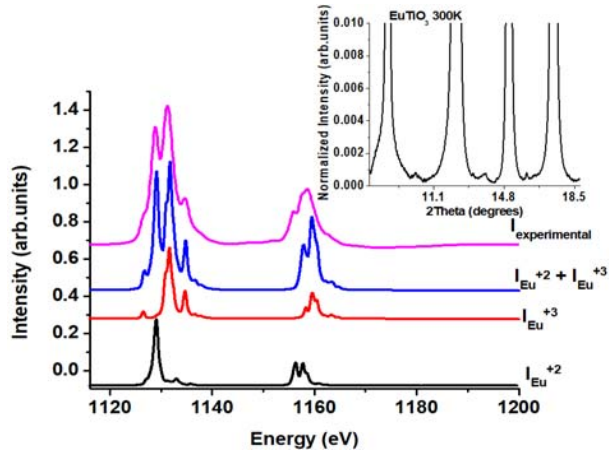

FIG. 8.

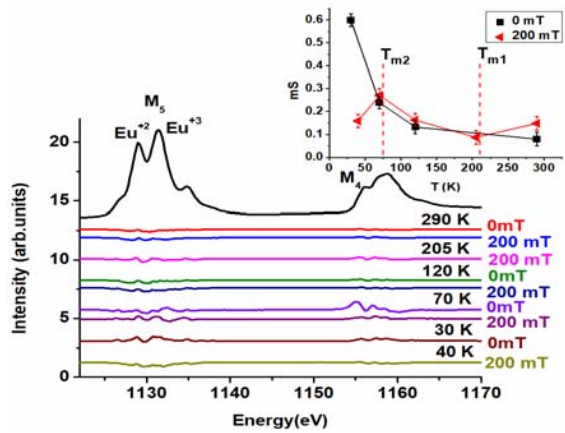

FIG. 9.